\documentclass[aps,pre,preprint,superscriptaddress,longbibliography]{revtex4-1}

\usepackage{graphicx,subfigure,subeqnarray,fancyhdr,amsmath,epstopdf,multirow,color, mathrsfs, amssymb}
\usepackage{ epsfig}
\usepackage{amsmath,amssymb,amsfonts}

\def\d{{\rm d}}

\begin{document}

\title{Helical propulsion in shear-thinning fluids}

\author{Saul G\'omez}
\affiliation{Instituto de Investigaciones en Materiales, Universidad Nacional Aut\'onoma de M\'exico, M\'exico}

\author{Francisco A. God\'inez}
\affiliation{Instituto de Ingenier\'ia, Universidad Nacional Aut\'onoma de M\'exico, M\'exico}

\author{Eric Lauga}
\email{e.lauga@damtp.cam.ac.uk}
\affiliation{Department of Applied Mathematics and Theoretical Physics, Centre for Mathematical Science,
  University of Cambridge, Wilberforce Road, Cambridge CB3 0WA, United Kingdom}
\author{Roberto Zenit}
\email{zenit@unam.mx}
\affiliation{Instituto de Investigaciones en Materiales, Universidad Nacional Aut\'onoma de M\'exico, M\'exico}

\date{\today}

\begin{abstract}
Swimming microorganisms often have to propel in complex, non-Newtonian fluids. We carry out experiments with  self-propelling helical  swimmers driven by an externally rotating magnetic field in shear-thinning, inelastic fluids. Similarly to swimming in a Newtonian fluid, we obtain for each fluid a locomotion speed which scales linearly with the rotation frequency of the swimmer, but with a prefactor which depends on the power index of the fluid. The fluid is seen to always increase the swimming speed of the helix, up to  50\% faster and thus the strongest of such type reported to date. The maximum relative increase for a fluid power index of around 0.6. Using simple scalings, we argue that the speed increase is not due to the local decrease of the flow viscosity around the helical filament but hypothesise instead that it originates from  confinement-like effect due to viscosity stratification around the swimmer.

\end{abstract}
\maketitle

\section{Introduction}
The physics of swimming microorganisms is a field of study which has long  supplied  different branches of natural sciences and engineering with an endless list of problems \citep{lighthill76}. Swimming cells provide, for example,  physicists with experimental models for active, out-of-equilibrium matter \citep{ramaswamy10,marchetti13} while  allowing engineers to draw inspirations to devise bio-mimetic designs \citep{nelson10}. In parallel, theorists have developed models allowing to interpret numerous phenomena from the natural world, such that  the  swimming of flagellated bacteria \citep{bergbook}, the role of fluid forces in reproduction \citep{fauci06}  and the physics of cilia-driven flows \citep{sleigh88,smith08}.

One of the areas of significant recent interest  aims to understand the effect of swimming in a complex fluid. In this case, work has primarily focused on attempting to capture locomotion either in real biological fluids, e.g.~mucus, or in model viscoelastic fluids. Physically, a non-Newtonian fluid provides a small  swimmer with at least two sources nonlinearities, namely the   shear-dependence of its viscosity and the emergence of elastic stresses \citep{morrison}. While both effects are typically present in any given biological configuration, they are physically very different and it is important to   separate their effects on locomotion in order to gain fundamental understanding of their impact on swimming microorganisms \citep{elfring2015theory}. Another source of non-linearity is the elasto-hydrodynamic coupling of appendage (e.g.~flagella) and flow. If the propulsion is the result of flapping flexible appendages then the swimming performance depends on the shape these appendices which, in turn, depends on the flow field around them~\citep{riley14,godinez2015}. This coupling  complicates therefore the understanding of each separate effect.

Locomotion in complex fluids has recently received a lot of attention, primarily in the form of theoretical investigations.  Small-amplitude analytical work showed how sheets and filaments are slowed down by elastic stresses
\citep{lauga07,FuPowersWolgemuth2007,FuWolgemuthPowers2009} while flexibility \citep{riley14}, the presence of  multiple traveling waves \citep{riley15} or viscosity stratification \citep{Man2015} can lead to enhanced locomotion. A similar framework was also developed for three-dimensional  geometries \citep{lauga_life,lauga14}. The impact of inelastic, shear-thinning stresses was  tackled using an asymptotic study of swimming sheets, showing that it was however a higher-order in most cases \citep{velezcordero13}, while the application to three-dimensional swimmers typically showed a non-monotonic decrease \citep{datt15}.

Numerical simulations have allowed to probe regimes not accessible by asymptotic work. To capture the role of elastic stresses, geometries tackled have included finite-size  sheets deforming at high amplitude \citep{teran2010,shelley13,becca14,li15}, spherical swimmers acting tangentially on the fluid \citep{lailai-pre,laipof1}, and infinite helices  \citep{spagnolie13}. The role of shear-thinning viscosity was addressed on two-dimensional \citep{loghin13} and spherically-symmetric swimmers  \citep{datt15} showing that a small enhancement was possible.

In contrast to the flurry of theoretical work, only a small number of experimental studies have probed in detail the impact of non-Newtonian stresses on locomotion, the majority of which focused on the role of elasticity. A helical filament driven in rotation in a viscoelastic fluid underwent slower force-free swimming than in the Newtonian limit in most cases, although a modest increase was possible at large helical amplitude \citep{liu2011}. On the other hand, rigid sheets \citep{dasgupta13} and  flexible filaments \citep{laugazenit13} undergoing planar waving motion were measured to swim faster in constant-viscosity elastic fluids. As an extreme case, reciprocal swimmers could be made to move due to elasticity \citep{keim12}.  The swimming of real flagellated bacteria in polymeric solutions was recently shown to be explained by Newtonian theory, while non-Newtonian effects at high molecular weight arose from the fact that  flagellar filaments are so thin that they essentially experience only the solvent viscosity \citep{martinez14}.

To date, only three experimental investigations have addressed the sole effect of shear-thinning viscosity on self-propulsion. \citet{dasgupta13} measured the propulsion speed of waving sheets embedded on freely-rotating cylinders. Elastic fluids with shear-thinning viscosities (carboxymethyl cellulose and polyethylene oxide) always lead to a decrease of the locomotion speed (for fixed characteristics of the wave) compared to the Newtonian limit. The only experimental studies of biological systems to date focused on the nematode {\it C. elegans} swimming in high-viscosity synthetic polymer solutions (xanthan gum) \citep{Arratia2013,Arratia2014}. While the change of fluid lead to changes in the flows produced by the swimming worm, no  change to the swimming kinematics was noticeable in the dilute limit, a result consistent with theoretical predictions in two dimensions \citep{velezcordero13}. In contrast, in concentrated system an increase of the swimming speed was reported  \citep{Arratia2013}.

\begin{figure}
\centering
\includegraphics[width=0.45\columnwidth]{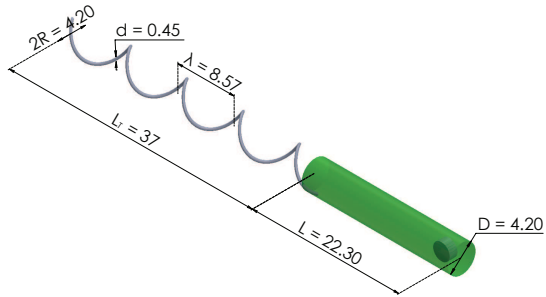}
\caption{Schematic view of the rigid synthetic swimmer: helical tail attached to a cylindrical head. The head shows the position of the permanent magnet. All dimensions are in mm.
}
\label{fig:swimmer}
\end{figure}

There is thus a need for  more systematic  studies addressing how a controlled variation of the rheological parameters of shear-thinning fluids impact well-defined swimming strategies. In this paper, we present the results for such a study. We measure experimentally the locomotion speed of macroscopic artificial swimmers composed of a rigid cylindrical body (with a small magnet enclosed at its tip) and a rigid helical filament rotated by an external magnetic field. These swimmers and their propulsion method are motivated by the helical locomotion of flagellated bacteria \citep{bergbook}.  Using inelastic fluids with shear-thinning viscosities  well-fitted to a power-law behaviour with a range of power indices from 0.47 to 1, we investigate how rheology impacts locomotion. Since the swimmers are propelled by a rigid helix, the kinematics of the propulsive element remain fixed and constant for all cases, simplifying one of the aspects of the problem.  Like in a Newtonian fluid, the locomotion speed of the swimmer is found to always scale linearly with the rotation frequency of the magnetic field, but with a prefactor which (i) is a function of the power index of the fluid; (ii) is always above the Newtonian prefactor, indicating enhanced locomotion; (ii) is maximum for a fluid power index of around 0.6. We  argue that this  increase above the Newtonian result is not due directly to the local decrease of the  viscosity around the helical filament but  instead that it originates from the recently-proposed confinement-like effect due to viscosity stratification around the swimmer \citep{li15,Man2015}.

\section{Experimental setup}

\subsection{Swimmer and actuation}

All experiments in this paper were conducted using the magnetic setup developed by \citet{godinez12}. The swimmer is  illustrated in Fig.~\ref{fig:swimmer}. A rotating magnetic field, generated by a Helmholtz coil pair, is used to actuate a small synthetic swimmer. The swimmer consists of a cylindrical plastic head of diameter $D=4.2$~mm and length  $L=22.3$~mm,  in which a permanent magnet is encased (Magcraft, models NSN0658). If the strength of the external magnetic field is sufficiently strong, the swimmer rotates at the same frequency. A rigid steel-wire helix of diameter $d=0.9$~mm is attached to the head which, as a result of its rotation, produces the thrust that propels the device. The diameter of the helix is $2R= 4.2$~mm, with a pitch angle $\theta=57^o$ and a length $L_T=37$~mm.  The swimmers were placed inside a  rectangular tank (160~mm $\times$ 100~mm $\times$ 100~mm) that fit into the region of uniform magnetic field inside the coils of approximately $(100$~mm)$^3$ in size where the test fluids were contained. For all the cases, the angular frequency of the rotating coils was below the step-out frequency~\citep{godinez12}; in other words, the swimmer rotates at the same rate as the external magnetic field.

\subsection{Test fluids and rheology}

\begin{figure}
\centering
\includegraphics[width=0.43\textwidth]{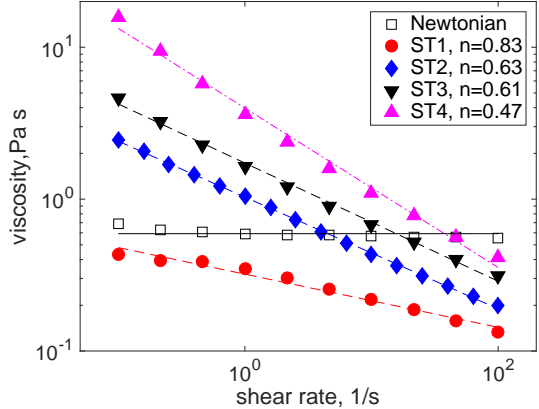}
\caption{Dynamic viscosity (in Pa.s) as a function of shear rate (1/s) for all the test fluids in steady  shear configuration. The lines indicate the fit to a power low model, Eq.~\eqref{PL}, for $0.1<\dot{\gamma}<100$ s$^{-1}$.}
\label{fig:rheology}
\end{figure}

Two types of fluids were fabricated, tested, and used: one Newtonian reference fluid and four shear thinning fluids with negligible viscoelasticity.  The rheological measurements were conducted using a rheometer (TA Instruments AR1000N) with a cone-plate geometry (60~mm, 2$^o$, 65~$\mu$m gap). We plot in Fig.~\ref{fig:rheology} the dynamic viscosity of the fluids under steady shear, while the physical properties of the solutions and their compositions are shown in Table~\ref{rheologicalproperties}.   Clearly, all  four non-Newtonian fluids display a power-law behavior in steady shear rates, $\dot\gamma$, within $0.1 < \dot\gamma < 100$ s$^{-1}$.
Their rheological behavior were closely fitted to a power-law model
\begin{equation}\label{PL}
\mu =m |\dot\gamma|^{n-1},
\end{equation}
where $m$ and $n$ are the consistency and power indices of the fluids, respectively, whose values are tabulated in Table~\ref{rheologicalproperties}. Furthermore,
within this range of shear rates, the rheometer did not register any measurable normal stress difference, and at least $N_1< 1$ Pa. To confirm negligible viscoelasticity, we also conducted oscillatory rheological tests, following the scheme proposed by \citet{velez-cordero11}; in all cases the storage modulus was smaller that the loss modulus. Furthermore, the relaxation time was estimated of order 10$^{-3}$ s; considering the characteristic shear rates in our experiments (see below), the viscoelastic effects are deemed to be negligible.  The Newtonian fluid, a glucose-water mixture,  was fabricated {a posteriori} in order to have a shear viscosity of the same order as that of the shear thinning fluids (see Table~\ref{rheologicalproperties}).

\begin{table*}[t]
  \centering
  \begin{tabular}{ c | c | c | c | c | c }\hline\hline
    Fluid & Composition & $n$ & $m$ & $\rho$ & Re$_{max}$\\
     & & (-) & Pa s$^n$& kg/m$^3$ \\ \hline
       N, ({$\square$}) & G/W, 91/9 & 0.99 & 0.662 & 1349 & 0.06 \\
       ST1, ({\color{red}$\bullet$}) & EG/C/TEA, 97.91/0.06/2.03 & 0.83 & 0.321 & 1116 & 0.18 \\
     \quad  ST2, ({\color{blue}$\blacklozenge$})\quad\quad  & \quad EG/C/TEA, 99.88/0.10/0.02 \quad&\quad 0.63\quad\quad & \quad1.031\quad\quad & \quad1113\quad\quad & 0.11\\
       ST3, ($\blacktriangledown$) & EG/C, 99.90/0.10 & 0.61 & 1.729 & 1113 & 0.12\\
       ST4, ({\color{magenta}$\blacktriangle$}) & EG/C/TEA, 99.27/0.70/0.03 & 0.47 & 3.979 & 1110 & 0.12\\
       \hline\hline
  \end{tabular}
  \caption{Physical properties of the all fluids tested in this investigation. The amount of each ingredient to make the fluids (water (W), glucose (G), ethylene glycol (EG), Carbopol (C) and triethyiamine (TEA)) is indicated in percentage by weight.}
  \label{rheologicalproperties}
\end{table*}

Since the viscosity of these fluids is dependent on the  rate of deformation, it is important to estimate  the  characteristic values of the shear rates around the swimmer. If $U$ denotes the linear velocity of the swimmer, $\omega$ its rotation rate  and $D$ the typical size of the head,  we can  estimate the characteristic  values of the shear rate due to translation and rotation of the head, namely  $\dot\gamma_{H, trans} = U/D$ and $\dot\gamma_{H,rot} = \omega$. Near the helical tail of thickness $d$, the estimate for shear rate in translation is  $\dot\gamma_{T,trans} = U /d$ while in rotation  it becomes $\dot\gamma_{T,rot} = \omega D/d$ since the filament moves with a typical velocity $\omega D$ through the fluid. For all cases the  shear rates range from 0.04 to 9.88 s$^{-1}$, which are within the range in which the test fluids have a clear power-law behaviour.

To validate our experimental technique and to ensure that there are no other effects, a set of additional measurements were conducted using Newtonian fluids with different viscosities, ranging from 0.1 to 4.5 Pa. The  speed of the swimmer in these fluids was measured to be identical in all cases (data not shown). Furthermore, the Reynolds number (calculated as to account for the shear dependent viscosity) is shown in Table \ref{rheologicalproperties}. For all cases $Re<0.18$, and therefore inertial effects are negligible.

\section{Results and discussion}
\subsection{Experimental results}

For each fluid, we measure the free-swimming speed of the device as a function of the angular frequency of the rotating magnetic field, all other parameters being kept fixed. No significant wobbling was observed and swimmers propelled essentially along straight lines, as can be seen in the movies provided in Supplementary Material. The maximum wobbling angle, $\theta$, was measured to be 1$\pm$ 0.1 degree, leading to negligible differences between body-frame and lab-frame swimming velocities.

\begin{figure}
\centering
\includegraphics[width=0.38\textwidth]{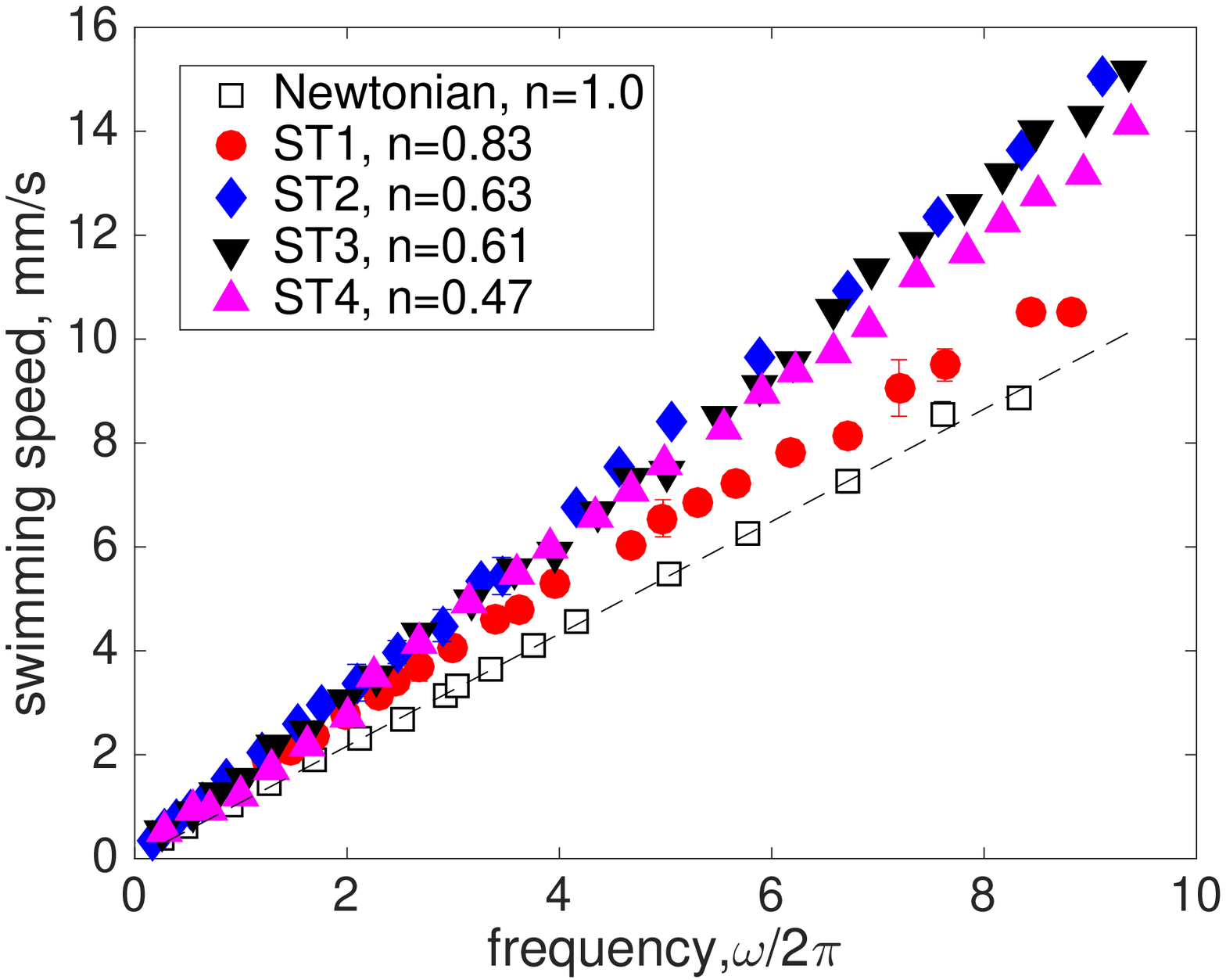}\quad\quad
\includegraphics[width=0.4\textwidth]{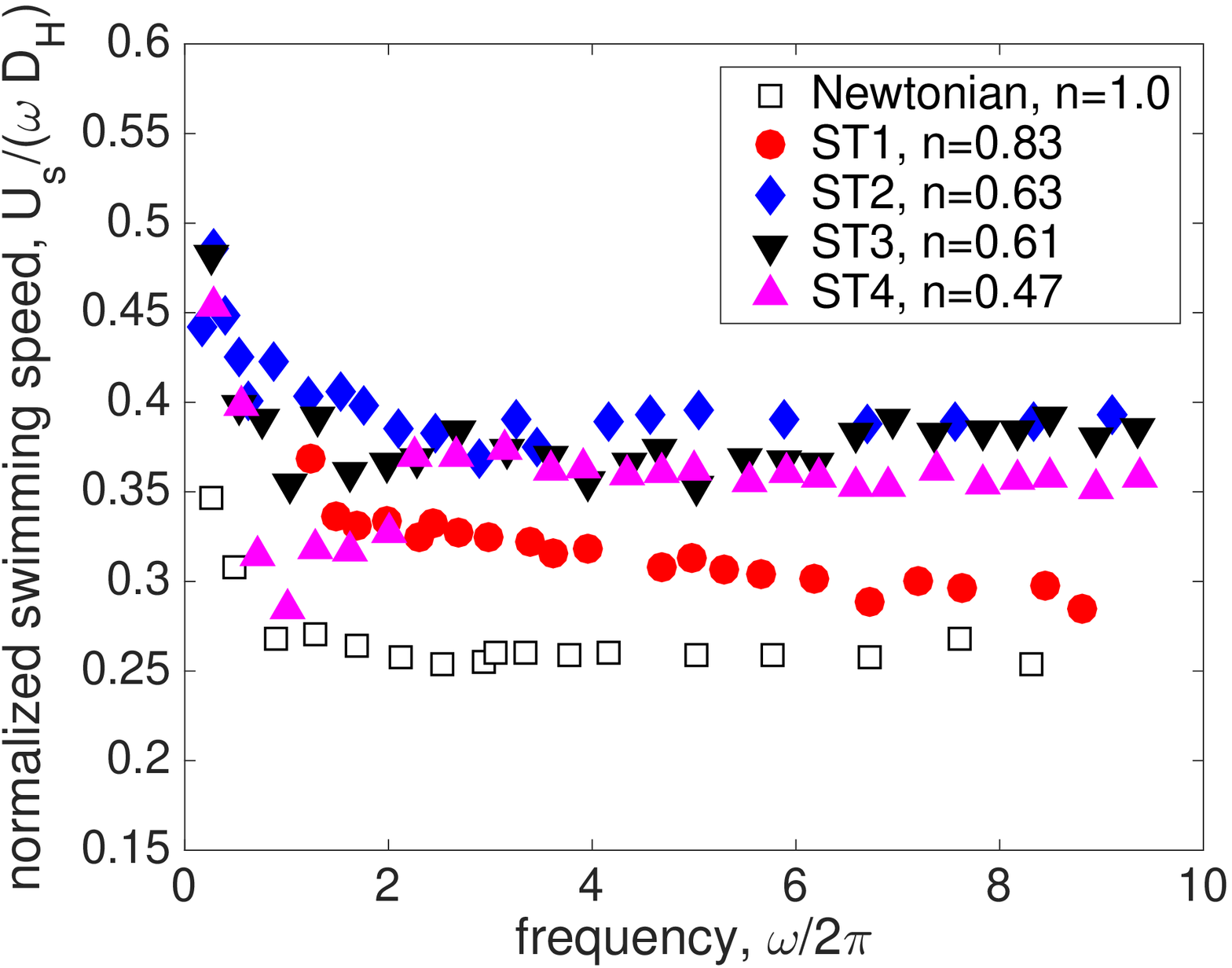}
\caption{
Left: Swimming speed,  $U$ (mm/s), as a function of frequency, $\omega/2\pi$ (1/s);
Right: nondimensionalised speed , $U/(\omega D)$, as a function of the rotational frequency, $\omega/2\pi$ (1/s), where $D$ is the diameter of the helix.}
\label{fig:velocity}
\end{figure}

The raw experimental results are plotted in Fig.~\ref{fig:velocity} (left)  for Newtonian (empty squares) and all four shear-thinning  fluids (filled symbols). Each data point shows the mean measured speed over four repeated experiments, and has error bars equal or  smaller than the one shown in the figure. As can be seen, the swimming speed scales linearly with the angular frequency, and is always above that obtained in the Newtonian case (empty squares).

In order to further validate this apparent linear dependence, we  plot in Fig.~\ref{fig:velocity} (right) the swimming velocity nondimensionalised  by the diameter of the swimmer and  the angular frequency. Except at very small frequencies  (where the uncertainty for small rotational speeds is larger), we observe an almost constant normalised velocity, function solely of the rheological properties of the fluid. Recall that in all cases the same swimmer is used.

\begin{figure}
\centering
\includegraphics[width=0.4\textwidth]{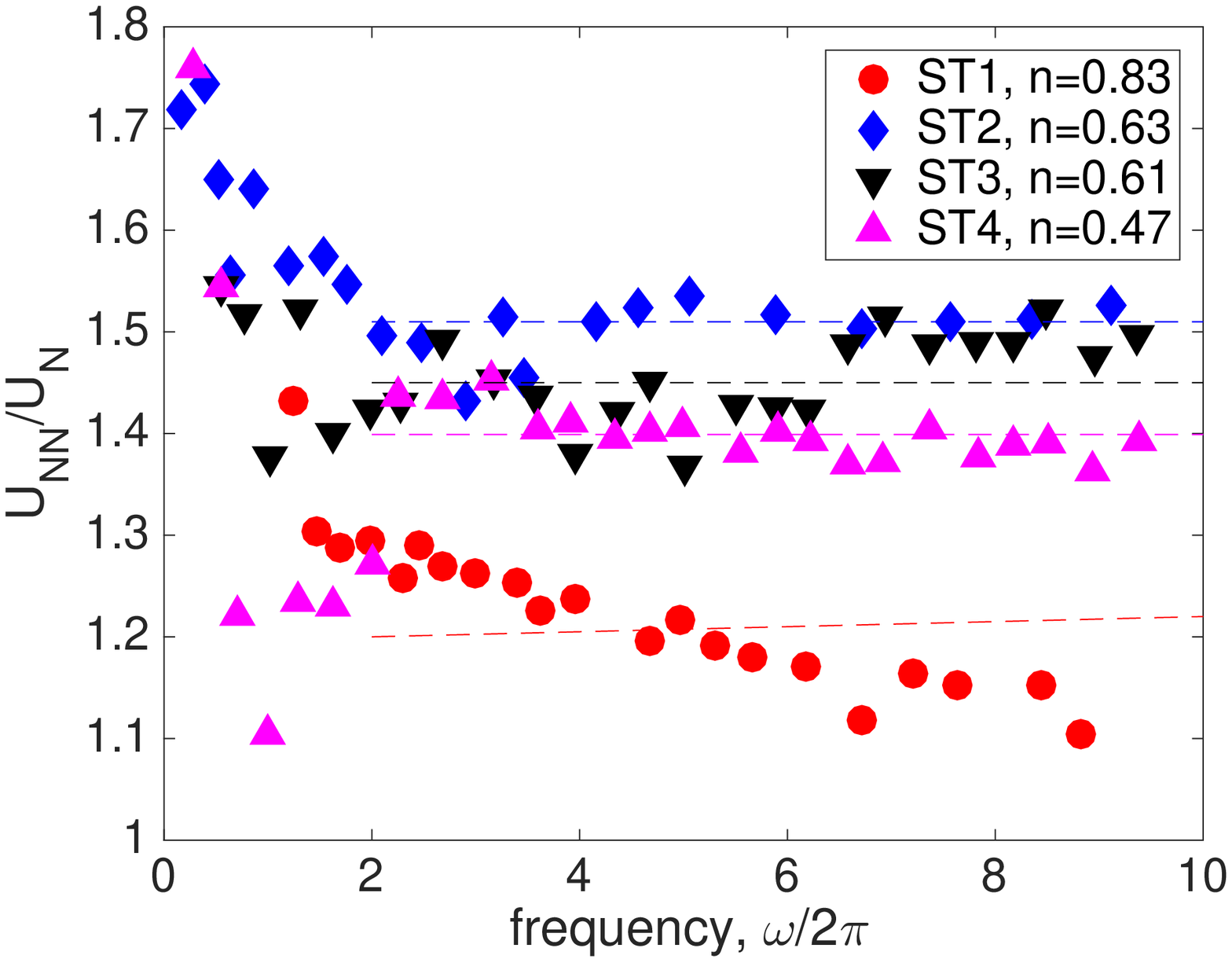}
\quad\quad
\includegraphics[width=0.4\textwidth]{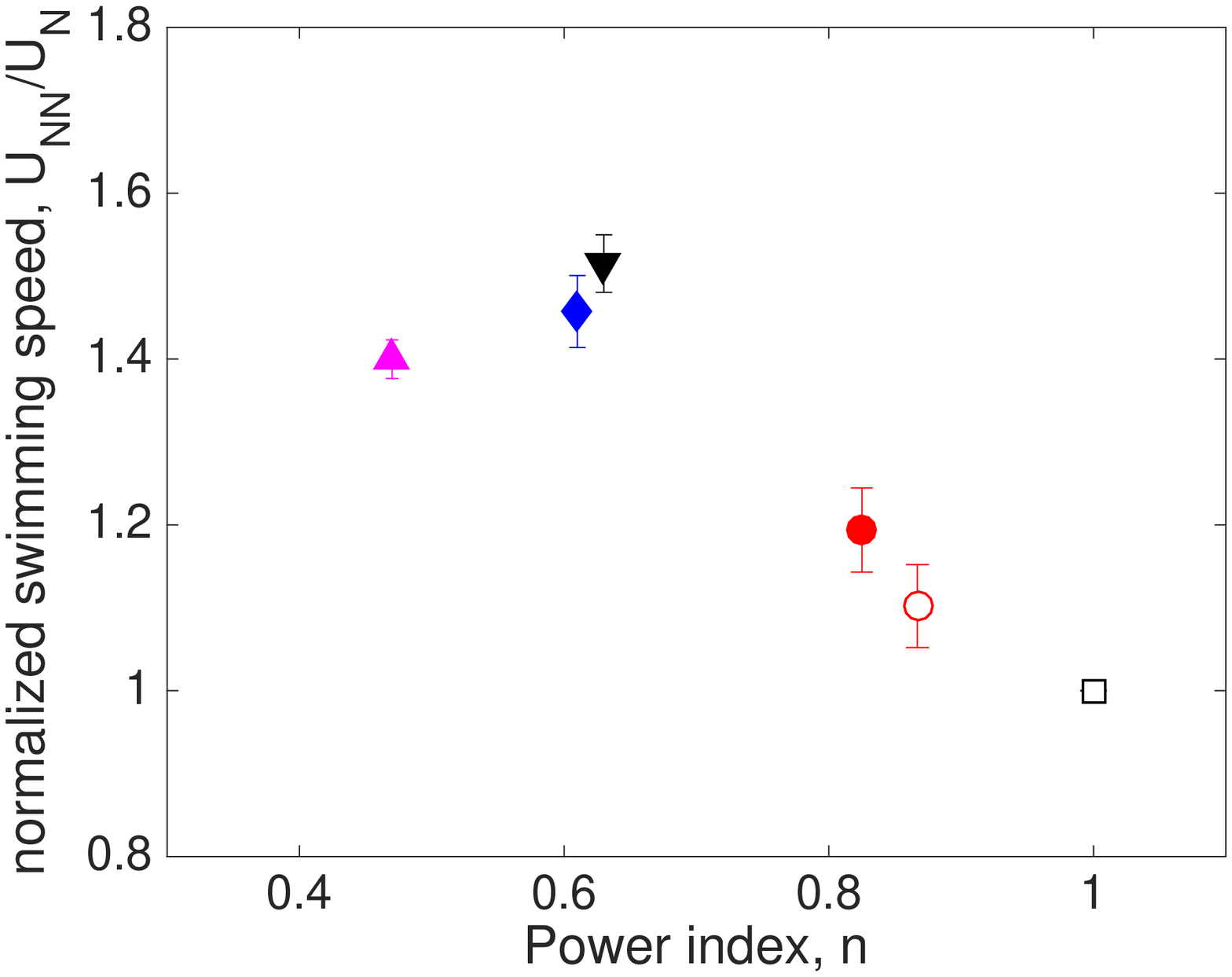}
\caption{Left: Ratio between the  swimming speed in a non-Newtonian fluid, $U_{NN}$,  and the Newtonian one, $U_N$, as a function of frequency of rotating magnetic field;
Right: Mean enhancement ratio, $\langle U_{NN}/U_N\rangle$, as a function of the fluid power index, $n$. }
\label{fig:velocity2}
\end{figure}

Since the swimming speeds in both Newtonian and shear-thinning fluids  scale approximately linearly with frequency, this suggests that their ratio will be constant. This is confirmed in Fig.~\ref{fig:velocity2} (left)  where we plot ratio between the non-Newtonian swimming speed, $U_{NN}$, and the Newtonian value, $U_N$,   as function of the angular frequency of the external field. Note that since the data in Fig.~\ref{fig:velocity} are not all obtained for the same frequency, we first fit the Newtonian data to a straight line (indicated by the dashed line in Fig.~\ref{fig:velocity}) and then divide the mean non-Newtonian results by this fitted line. Swimming enhancements of up to  50\% are obtained,  the strongest  reported to date experimentally.

The results in Fig.~\ref{fig:velocity2} (left) confirm a systematic, and  roughly constant, enhancement of speed above the Newtonian results. We then compute the mean value of this speed enhancement, as shown by the dashed lines, and plot  its dependence on the power index of the fluid in Fig.~\ref{fig:velocity2} (right). Strikingly, and beyond experimental errors (error bars are indicated on the figure) the enhancement is non-monotonic: the fluids with a power index of $n\approx 0.6$ appear to lead to the largest increase in swimming above the Newtonian value.

Note that in Fig.~\ref{fig:velocity2} (right), there is an additional data point at $n=0.87$ (empty red circle). This data corresponds to additional tests conducted with the ST1 fluid (filled red circles), but after two months. It is well known  that the rheological properties to these fluids evolve in time \citep{brennen1967}. Therefore, the fluid was characterised again. Both consistency coefficient and power index had changed slightly ($n=0.87, m=0.405 $Pa s$^n$). The results obtained with this fluid are consistent with the rest of the data.

\subsection{Discussion}

Based on our experimental results, it is clear that one  fundamental question needs to be answered, namely why does a change of the fluid from Newtonian to non-Newtonian lead to an increase in the swimming speed?

\begin{figure}
\centering
\includegraphics[width=0.4\textwidth]{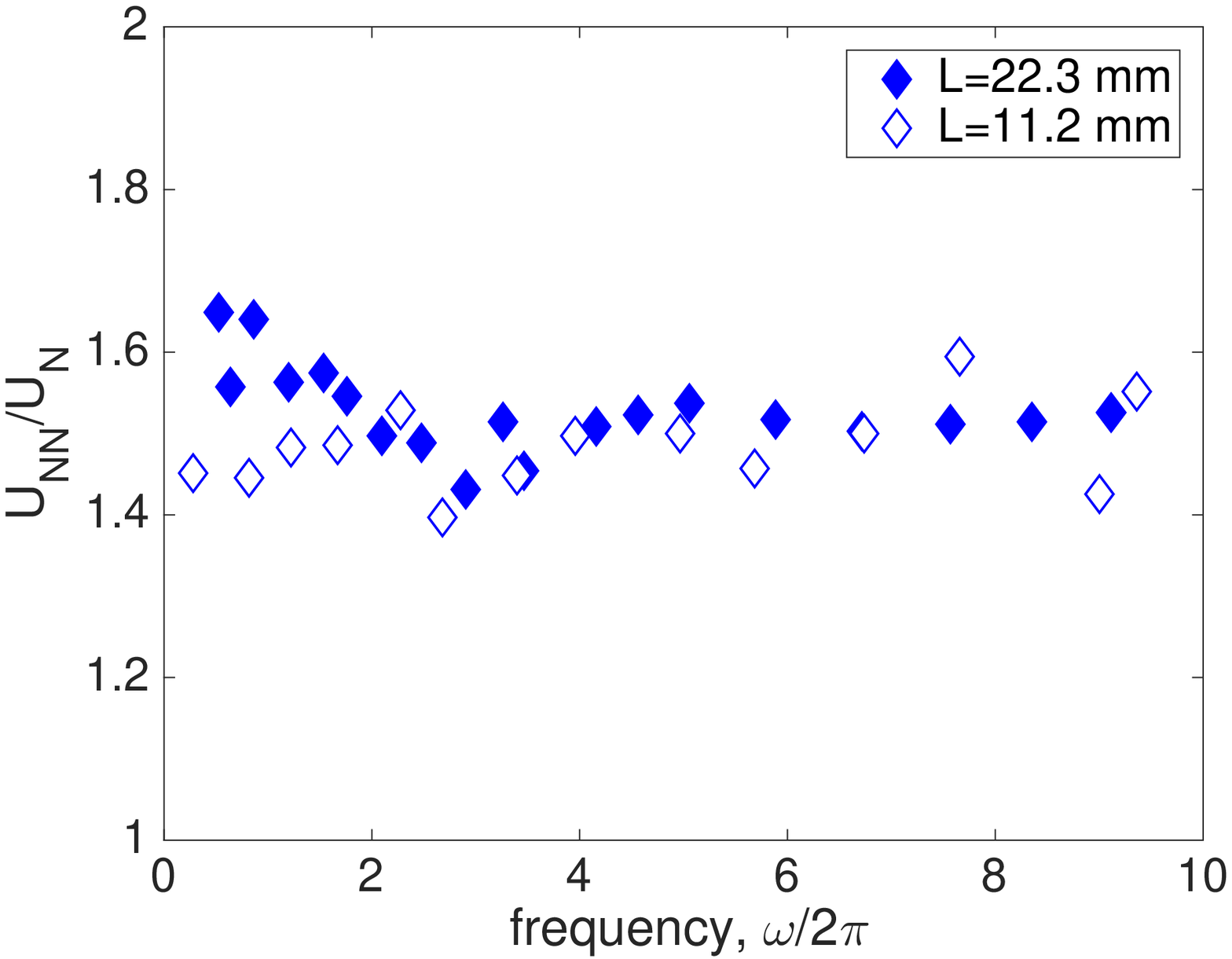}
\quad\quad
\includegraphics[width=0.4\textwidth]{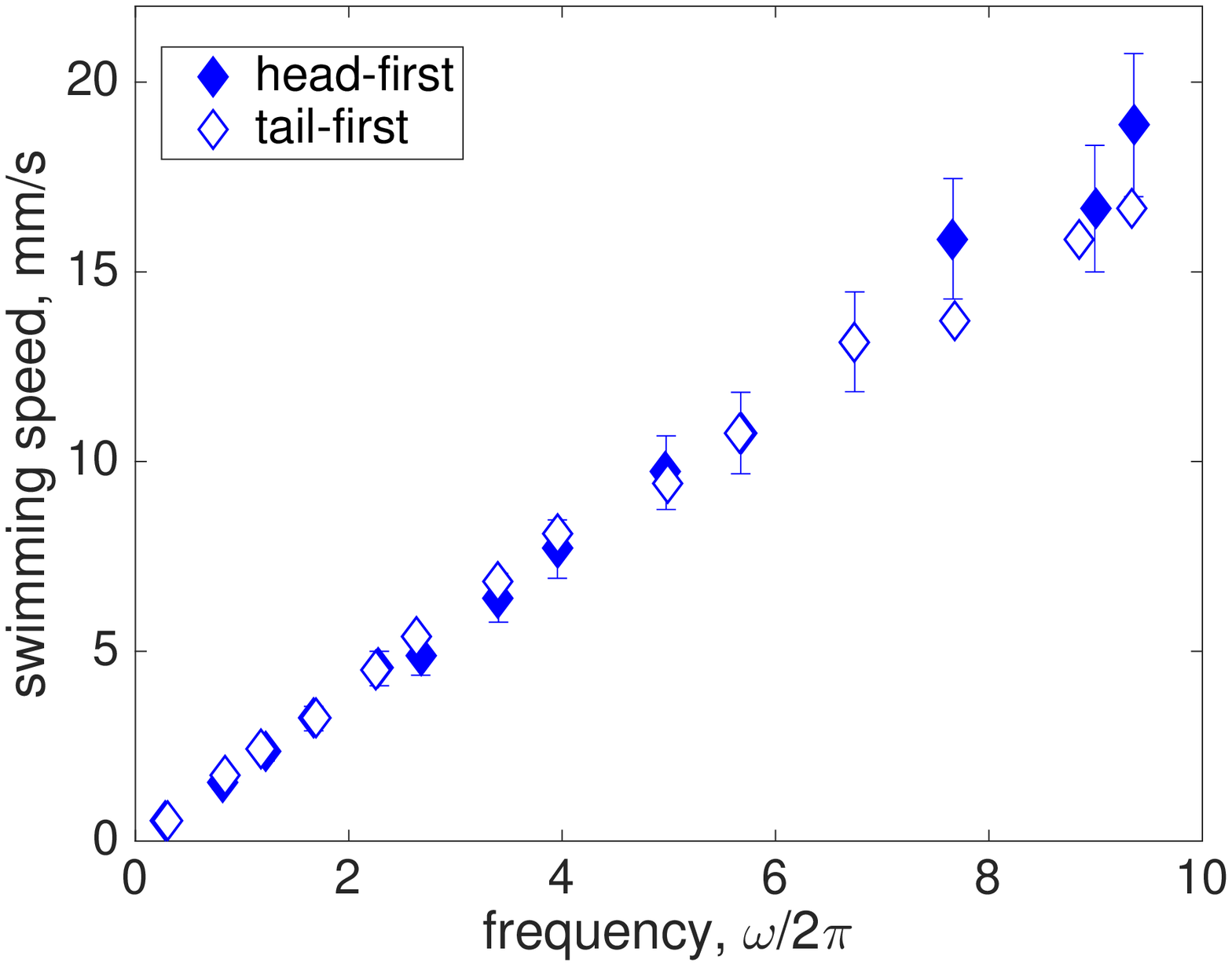}
\caption{Left: Ratio between the  swimming speeds, $U_{NN}/U_N$, as a function of frequency of rotating magnetic field for swimmers with two different head lengths, $L$; Right: Swimming speed,  $U$ (mm/s), as a function of frequency, $\omega/2\pi$ (1/s) for two directions of motion (head-first and tail-first). For both plots the ST2 fluid  was used (see Table \ref{rheologicalproperties}).}
\label{fig:velocity3}
\end{figure}

Four  different physical mechanisms  can be brought forward to explain the increase in swimming speed. The first one is the role played by the  head of the swimmer. The rotation of the head leads to a decrease of the viscosity of the fluid surrounding it, and thus a decrease of its drag. For an unchanged  propulsion force, this would lead to a faster  swimming speed, similarly to recent experimental observations in sedimentation of rotating spheres \citep{godinez14}. To rule out this effect, we performed additional experiments with swimmers with half-as-long heads. It was found that the ratio between non-Newtonian an Newtonian speeds remain unchanged despite the difference in head length. One of these results is shown in Fig. \ref{fig:velocity3} (left).

A second hypothesis could be the role of a non-Newtonian wake. As the swimmer progresses through the fluid head-first, the head lowers the viscosity of the fluid. A cloud of low-viscosity fluid is dragged behind the head that could affect the creation of thrust by the helical filament. This effect is likely to be small for inelastic fluids with no memory, but in order to rule it out, we performed experiments where we rotated the magnetic field in the opposite direction, leading to helix-first swimming. The results were unchanged and the same speeds were measured for all cases. Results of a typical experiment showing this behaviour are displayed in Fig. \ref{fig:velocity3} (right).  Note that since the typical shear rate  near the helical filament is actually larger than that around the head, an even lower viscosity wake could potentially  be induced by the tail rather that the head in the reverse motion. Since the data in Fig.  \ref{fig:velocity3} (right) does not show any directional preference, this argument clearly does not explain a faster swimming in shear thinning fluids.

A third route to a change in the swimming speed could come from the difference in local viscosity near the swimmer head (typical value denoted $\mu_H$) and helical tail ($\mu_T$) \citep{martinez14}. To address this, we consider a  simple Newtonian-like model.  The swimming speed of the device comes from the balance between thrust created by the helix, $T$,  and drag from both the helix ($D_T$) and the body ($D_H$), leading to an overall force-free motion. Assuming a locally Newtonian fluid with viscosity $\mu_T$, the thrust and drag created/experienced  by the helix of radius $R$ as it rotates with frequency $\omega$ and translates at speed $U$ scale as
 \begin{equation}\label{tail}
Tr\sim \mu_T \omega R^2 f,\quad D_T\sim \mu_T U R g,
\end{equation} where $f$ and $g$ are dimensionless functions of geometry of tail. Similarly, the head drag  scales as
\begin{equation}\label{head}
D_H\sim \mu_H U D h ,
\end{equation}
where $D$ is the head diameter and $h$ a dimensionless function of the body geometry.
Balancing Eq.~\eqref{tail} and Eq.~\eqref{head} as  $D_T+D_H\sim Tr$ leads to  the scaling
\begin{equation}
\mu_T U R g + \mu_H U D h  \sim  \mu_T \omega R^2 f,
\end{equation}
and thus a swimming speed approximately given by
\begin{equation}
U  \sim\frac{   \omega R f}{\displaystyle\left(g + \frac{\mu_H}{\mu_T} \frac{D}{R} h\right)}\cdot
\end{equation}
The problem then boils to understanding how the ratio of viscosities is expected to vary with $n$. For fixed geometry and rotation rate, since the fluid follows a power-law rheology, the ratio of viscosities is on the order of the ratio of shear rates.
\begin{equation}
\frac{\mu_H}{  \mu_T}\sim \left(\frac{\dot\gamma_H}{\dot \gamma_T}\right)^{n-1}.
\end{equation}
  However, since there is a large difference in diameter between the body and the helical filament ($d\ll D$), we have    $\dot\gamma_H \ll \dot \gamma_T$, and thus the ratio of viscosities is a small number to the power $n-1$ which decreases with  $n$, i.e.
\begin{equation}
\frac{\d }{\d n} \left(\frac{\mu_H}{\mu_T}\right) < 0.
\end{equation}
A decrease  in $n$ will thus increase the viscosity ratio and decrease the swimming speed. This is the exact opposite of what is seen in our experiments, and therefore another physical mechanism is at the origin of the speed increase.

We hypothesise in this paper that the enhancement of locomotion  is due to the gradients in  the fluid viscosity which results from the spatial gradients in the shear rates. In other words, the simple Newtonian scaling above is not correct because it ignores the fact that swimming in a viscosity gradient is akin to swimming under (soft) confinement. An increase of the swimming speed resulting from confinement was first discussed by \citet{katz74} for the case of Newtonian fluids; the subject has recently been studied numerically \citep{liu2014}. Consequently, in Eqs.~\eqref{tail} and \eqref{head}, the prefactor is not a constant but depends also on the gradients in viscosity. This conjecture is consistent with two recent studies which have shown  this confinement to be responsible for significant swimming enhancement in two and three dimensions \citep{Man2015,li15}. In their numerical study  \cite{li15} considered the locomotion of a waving sheet; they showed that the field of viscosity around the sheet for the cases where swimming enhancement was observed displayed a well defined corridor of low viscosity fluid confined by a high viscosity region. \citet{Man2015} analytically showed that a waving sheet or a three-dimensional filament would swim faster when it is surrounded by a  higher-viscosity fluid, arguing that  the viscosity gradient affects the ratio of normal to tangential forces. In our case, the shear induced by the rotation of the head and the tail gives the fluid in the vicinity of the swimmer  a smaller value of viscosity than that at larger distances.  As a simple mathematical model for this, consider the rotation of a cylinder of radius $a$ immersed in a fluid with power-law behaviour; this could represent either the head  of the swimmer (diameter $D$) or its tail  (diameter $d$). Solving for the two-dimensional Cauchy equation for the shear stress, and then inverting the equation to derive the velocity field allows to compute the effective local viscosity, $\mu_{\rm eff}$, analytically and we obtain
\begin{equation}
  \mu_{\rm eff}= m \left[\frac{2\omega}{n}\left(\frac{a}{r}\right)^{2/n}\right]^{n-1}.
  \label{eff_mu}
\end{equation}
Clearly, the local viscosity in the fluid is spatially dependent, with low viscosity near the cylinder and increasing away from it, in a manner which depends on both the shear-thinning properties of the fluids ($m$ and $n$) and on the rotational speed $\omega$. With this simple approach, how strong are the gradients in the viscosity? Using Eq.~\eqref{eff_mu} we can compute the value of the viscosity gradient near the cylinder and obtain
\begin{equation}\label{visc_grad}
\frac{\partial \mu_{\rm eff}}{\partial r}\bigg\vert_{r=a} =\frac{m}{a\omega}\frac{(1-n)(2\omega)^n}{n^n}\cdot
\end{equation}
Remarkably, the function in Eq.~\eqref{visc_grad} is not necessarily monotonic in $n$; it is always increasing for decreasing values of $n$ near 1, i.e.~for fluids slightly shear thinning. However, for some values of $\omega$, the viscosity gradient can reach a maximum value at a finite value of $n$ before decreasing when one decreases $n$ further. This simple model is thus  consistent with both the increase in $U_{NN}/U_N$ for $n\lesssim1$ and the observed maximum in the ratio  at an intermediate value $n\approx0.6$.  We note that, similarly, \citet{li15} found that if the ratio of viscosities between the inner and outer regions is too large the swimming enhancement is weakened.

\section{Conclusions}
In this paper, we have conducted experiments with helical swimmers with a fixed shape that self propel under the action of an external magnetic field in well characterized shear-thinning inelastic fluids. It was found that for all the cases considered the swimming speed scales linearly with the actuation frequency, in agreement with the prediction for Newtonian fluids. However, and most relevant, the swimming speed for a given frequency was always larger for the shear thinning fluid than that for the Newtonian case. The ratio $U_{NN}/U_N$ was found to be a function of the power law index $n$, increasing from the Newtonian case ($n=1$) to reach a maximum at $n\approx0.6$ to then decrease as $n$ decreased further.  Considering the different scenarios that could lead to an enhancement in the swimming speed, we ruled out the effect of the head, the effect of a wake and the effect of a contrast between the local viscosities of the head and the tail. The only argument consistent with our results (arguably obtained indirectly by discarding all other possible effects)  was the confinement-like effect.
 In a manner similar to that predicted for helices within solid confinement, and more recently for waving sheets  in shear thinning fluids \citep{li15} and waving sheets and filaments in a region of low viscosity surrounded by a high-viscosity fluid \citep{Man2015}, the swimming speed of the helical swimmer is larger than the unconfined case. In order to fully unravel the physical origin of enhanced swimming in a shear thinning fluid, a visualisation of the fluid structure  around the swimmer, using for example PIV, would allow to verify our hypothesis. We hope that our results will encourage future work along these lines.

\section*{Acknowledgments}
The help of C.~Bellevile with some of the experiments is gratefully acknowledged. We thank M.~Ramirez-Gilly for her assistance in the rheological characterization of the test fluids.  This work was funded in part by the European Union (CIG grant to E.L.). R.Z. acknowledges the financial support of the Moshinsky Foundation and the PAPIIT-DGAPA-UNAM program (grant no. IN101312).
\bibstyle{unsrt}
\bibliography{helical_swimmers}
\end{document}